\documentclass[12pt]{iopart}
\usepackage{epsfig}
\begin{document}
\title[Results from FOPI on strangeness production and propagation at SIS 
energies]{Results from FOPI on strangeness production and propagation at SIS 
energies}

\author{Anne Devismes
\footnote[3]{a.devismes@gsi.de}
 for the FOPI collaboration}

\address{GSI-KP1, Planckstrasse 1, 64291 Darmstadt, Germany}

\begin{abstract}
Heavy ion collisions at SIS energies (1-2 AGeV) offer an unique tool to probe 
the properties of hot and dense nuclear matter. In particular, the 
partial restoration of chiral symmetry is predicted to lead in this energy 
range to in-medium modifications of hadron properties. Strange particle 
production below or close to the threshold energy is a useful 
probe to investigate these in-medium effects. The FOPI collaboration has 
recently measured the production and the propagation of charged and neutral 
strange particles. The $\rm K^{+}$ production probability is investigated 
as a function of the system size at a beam energy of 1.5 AGeV. Results on 
$\rm K^{0}$ production in Ru+Ru collisions at 1.69 AGeV are presented, as 
well as $\rm K^{-}$/$\rm K^{+}$ ratio as a function of rapidity. 
In addition, the sideward flow of charged and neutral strange particles has 
been measured. Results are compared to predictions of transport 
calculations (BUU and IQMD). 
\end{abstract}




\section{Introduction}

For the last two decades, strangeness production has played a very 
important role in nuclear physics. Strangeness enhancement was proposed as 
a signature of the quark gluon plasma in the early 80's~\cite{raf82}. At the 
same time, lots of efforts have been devoted to the investigation of hadron 
properties in a hot and dense medium, both on the theoretical and 
experimental side, leading to the dropping mass scenario~\cite{bro91}.\\
Kaons are of particular interest due to the sensitivity of their properties 
and propagation to the state of the nuclear matter in which they have been 
produced. Their properties may reflect the spontaneous breaking of chiral 
symmetry and its restoration.\\
Due to the presence of a repulsive kaon-nucleon potential and an attractive 
antikaon-nucleon potential, kaon and antikaon masses vary differently as 
a function of the nuclear matter density. The kaon mass increases with the 
density whereas the antikaon mass decreases. As a consequence, it becomes 
energetically more difficult to produce a kaon and easier to produce an 
antikaon in a dense medium. This property can be observed by measuring kaon 
and antikaon yields in heavy ion collisions and comparing the results to 
theoretical predictions. Strangeness exchange reactions such as 
$\rm K^{-} N \Longleftrightarrow \pi \Lambda$ seem to play an important 
role in kaon production and have therefore to be taken into account.\\
The kaon-nucleon potential also influences kaon propagation which 
is measured in terms of flow~\cite{cro00}.\\
This contribution is organized as follows: the method and quality of 
particle identification with the FOPI detector is presented in the next 
section. Section 3 concerns strangeness production. $\rm K^{+}$ production 
has been measured as a function of the system size. $\rm K^{+}$ and 
$\rm K^{0}$ rapidity distributions are presented as well as 
$\rm K^{-}/K^{+}$ ratio. In section 4, results on the sideward flow of 
charged and neutral strange particles are summarized. 

\section{FOPI detector and particle identification}

\begin{figure}[hbt]
\vspace{0.5cm}
\centering\mbox{\epsfig{file=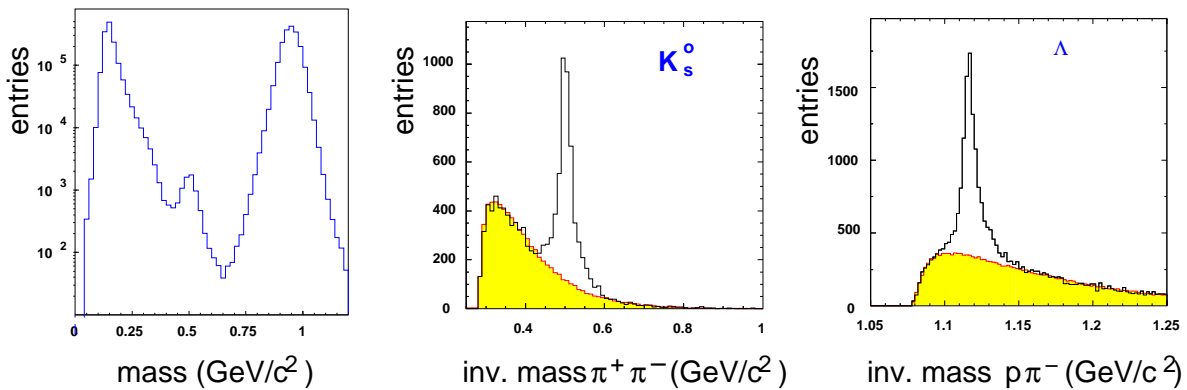, width=1.\textwidth}}
\vspace{2mm}
\caption{Mass distribution for Ru+Ru collisions at 1.69~AGeV (left) 
and $\pi^{+} \pi^{-}$ (middle) and p$\pi^{-}$ (right) invariant mass spectra. 
On the middle and right panels, the shaded area shows the combinatorial 
background.} 
\label{massedistri}
\end{figure}

The FOPI detector~\cite{gob93}, installed at SIS/GSI (Darmstadt) has a 
relatively large coverage of the phase space and is composed of several 
sub-detectors. The central part is placed in a solenoid providing a 
magnetic field of 0.6~T and consists of a drift chamber (CDC) and a 
barrel of plastic scintillators. The forward part consists of a second 
drift chamber (HELITRON) and a wall of plastic scintillators. The 
later provides a measurement of the charged particle multiplicity 
in the forward hemisphere, used for centrality selection.\\
The results presented in this contribution have been measured in the central 
part of the FOPI detector (CDC + BARREL). The measurement of the track 
curvature in CDC due to the magnetic field is combined with the energy loss 
of particles and provides a first mass identification (Bethe-Bloch) as well 
as a charge sign identification. A second mass identification is obtained 
when a track in the CDC can be matched to a hit in the BARREL that provides 
a time-of-flight measurement.\\ 
An upper momentum cut of 0.5~GeV/c is needed to properly separate charged 
kaons from pions and protons. A mass distribution is presented on 
figure~\ref{massedistri} (left panel) for Ru+Ru collisions at 1.69~AGeV.

\begin{figure}[hbt]
\vspace{0.5cm}
\centering\mbox{\epsfig{file=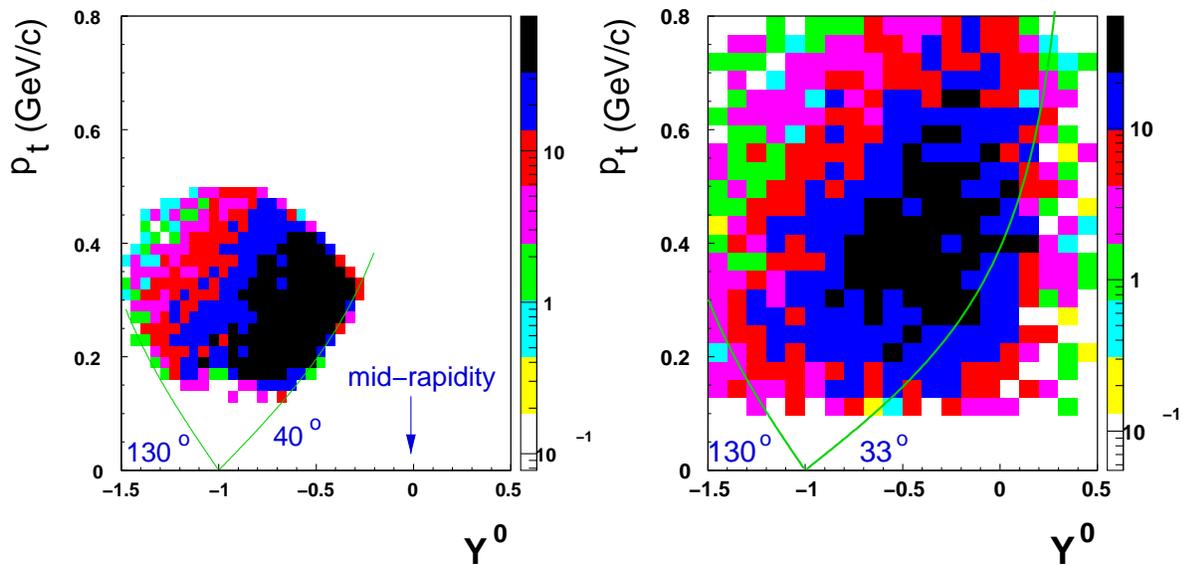, width=1.\textwidth}}
\vspace{2mm}
\caption{Acceptance in terms of $\rm p_{t}$ as a function of rapidity 
for $\rm K^{+}$ (right) and $\rm K^{0}$ (left). The rapidity is normalized 
to the beam rapidity.} 
\label{acc}
\end{figure}

Neutral kaons and $\Lambda$ are reconstructed from their decay into 
$\rm \pi^{+}$ $\rm \pi^{-}$ and p $\rm \pi^{-}$, respectively. 
Only the short component $\rm K^{0}_{S}$ with a c$\tau$ of 2.68~cm can be 
measured in the FOPI detector, the $\rm K^{0}_{L}$ decaying outside of 
the apparatus. Invariant mass spectra are shown on figure~\ref{massedistri} 
for $\rm K^{0}_{S}$ (middle panel) and $\Lambda$ (right panel). The dark 
area represents the combinatorial background obtained from event mixing.\\
The results presented in what follows have been corrected for matching and cut 
efficiencies using protons with the same momentum for $\rm K^{+}$ corrections 
and a detailed GEANT-based simulation for neutral strange particles.\\
Figure~\ref{acc} shows the acceptance for $\rm K^{+}$ (left panel) and 
$\rm K^{0}$ (right panel) in terms of transverse momentum as a function of 
rapidity. For charged kaons, the acceptance is limited by the polar angle 
coverage of the central barrel 
($\rm 40^{\circ} < \theta_{lab} < 130^{\circ}$) and by the upper momentum cut 
of 0.5~GeV/c needed to properly identify kaons. In addition, a transverse 
momentum of at least 0.1~GeV/c is needed for the particles to reach the 
time-of-flight barrel. In the case of neutral particles, the acceptance is 
larger due to the kinematics of the decay and covers the whole backward 
hemisphere.

\section{Strangeness production}

As already mentioned, strangeness production is connected to fundamental 
aspects of nuclear physics and appears as a promising tool to probe the 
modification of hadron properties in-medium. However, those properties 
may depend as well on the nuclear equation of state. For this reason, it is 
necessary to perform systematic analysis of various systems and to compare 
those results to theoretical predictions. This has been investigated by 
the KaoS collaboration~\cite{stu01}.

\subsection{Dependence of $\rm K^{+}$ production on the system size}

\begin{figure}[hbt]
\vspace{0.5cm}
\centering\mbox{\epsfig{file=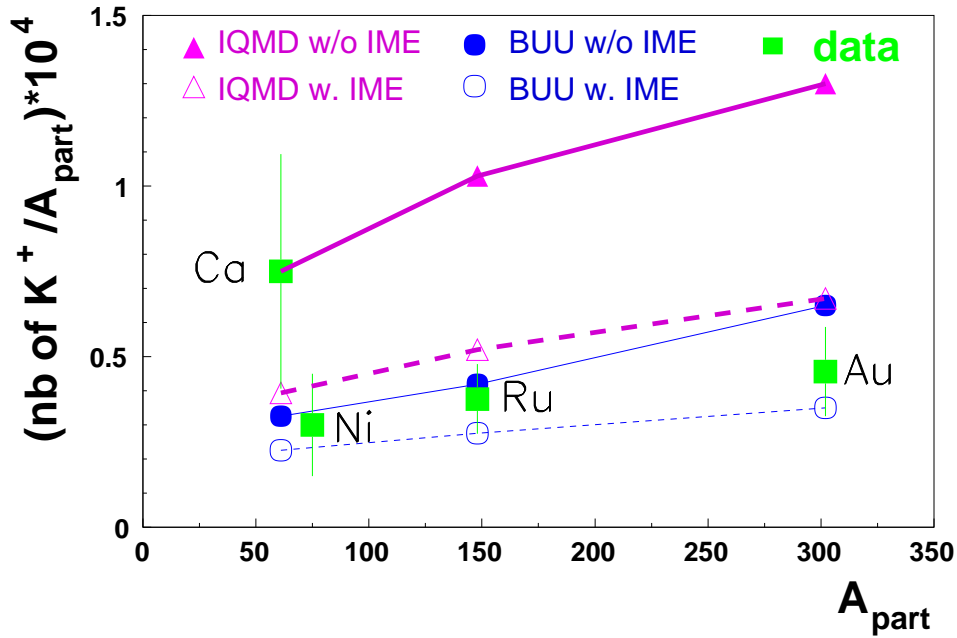, width=.8\textwidth}}
\vspace{2mm}
\caption{${\rm K}^+$ production as a function of the system size at a beam 
energy of 1.5~AGeV. Data are shown by the squares. Triangles (circles) show 
the predictions of the IQMD (BUU) model without (solid symbols) and 
with (open symbols) in-medium effects.}
\label{kpapart} 
\end{figure}

The FOPI collaboration has recently investigated the $\rm K^{+}$ production 
as a function of the system size at a beam energy of 1.5~AGeV. The number of 
charged kaons has been measured in the acceptance of the FOPI central part 
in central Ca+Ca, Ru+Ru and Au+Au collisions~\cite{dev01}. An older point for 
Ni+Ni collisions has been included~\cite{bes97}. Data are corrected for cut 
and matching efficiencies. Results are presented on figure~\ref{kpapart} 
(squares). The calcium point has been obtained with a very poor statistics. 
A slight increase from the Ni system to the Au system is observed. This 
increase is predicted by transport calculations 
(BUU (Boltzmann-Uehling-Uhlenbeck)~\cite{cas99} and IQMD 
(Isopin Quantum Molecular Dynamic)~\cite{har92}). The data seem to favor the 
version with in-medium effects for the IQMD model (open triangles). It is hard 
to draw any conclusion for the BUU model.

\subsection{$K^{+}$ and $K^{0}$ rapidity distribution}

Rapidity distributions are obtained by fitting transverse mass spectra 
in narrow rapidity windows by a Boltzmann function. The integral of the 
fit function between $m_{0}$ and $\infty$ gives the yield. Results for 
central Ru+Ru collisions at 1.69~AGeV are presented on figure 4. 
$\rm K^{+}$ ($\rm K^{0}$) data are shown by the squares (dots). 
$\rm K^{0}$ yields have been multiplied by a factor of 2 to account for 
the $\rm K^{0}_{L}$ component that is not measured with our apparatus. 
Charged and neutral kaon yields agree with each other within error bars.\\
Solid (dashed) lines show the results of transport calculations without 
(with) in-medium effects for BUU (thin lines) and IQMD  (thick lines). 
The largest effect appears in the mid-rapidity region where the densities 
are the highest. This shows the importance of measuring strangeness 
production close to mid-rapidity. For the first time, the FOPI collaboration 
presents kaon data in this region of phase space. \\ 
The IQMD model overestimates the data with the two different scenarios 
whereas the version of BUU without in-medium effects agrees with the data 
over the whole rapidity range.\\
The differences between the models is due to different elementary 
cross sections used as input. It has been checked that the same cross 
sections lead to the same results with both models~\cite{har01}. 

\begin{figure}[hbt]
\includegraphics{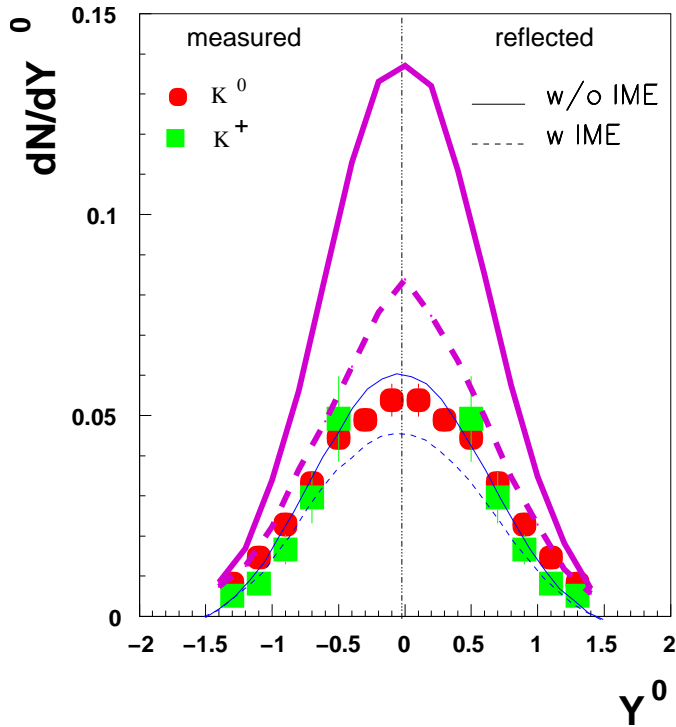}
\vspace{1.8mm}
\vspace{2cm}
\hspace{8cm}
\begin{minipage}{6.8cm}
\baselineskip=12pt
{\begin{small}
\caption{Rapidity distributions for $K^{+}$ (squares) and $K^{0}$ 
(dots) in Ru+Ru collisions at 1.69~AGeV. The thick lines show the IQMD 
predictions and the thin lines show the BUU calculations without 
(solid lines) and with (dashed lines) in-medium effects.}
\vspace{3.6cm}
\end{small}}
\end{minipage} 
\label{dndykpk0} 
\end{figure}

\subsection{$\rm K^{-}$/$\rm K^{+}$ ratio}

\begin{figure}[hbt]
\vspace{0.5cm}
\centering\mbox{\epsfig{file=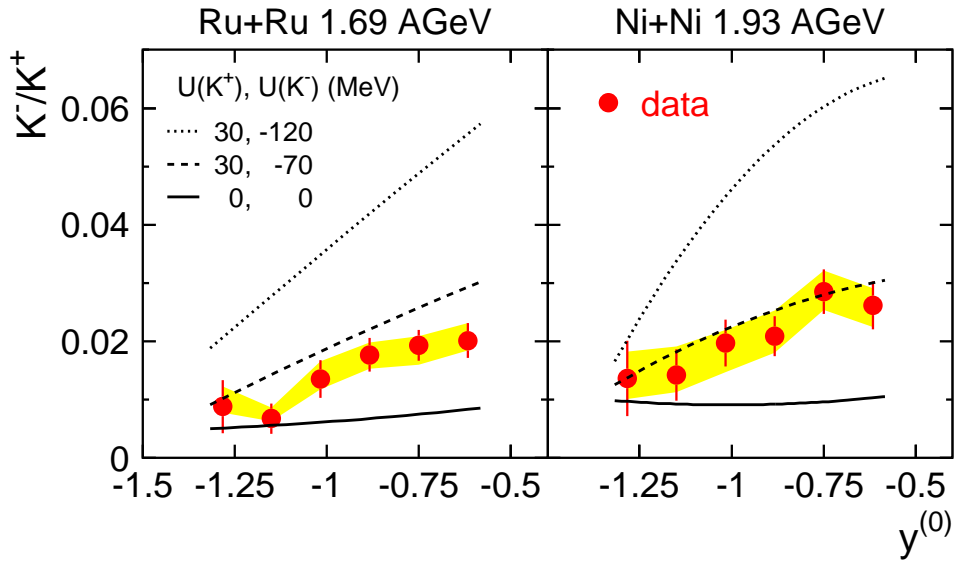, width=.8\textwidth}}
\vspace{2mm}
\caption{$\rm K^{-}$/$\rm K^{+}$ ratio as a function of the rapidity for 
Ru+Ru collisions at 1.69~AGeV (left panel) and Ni+Ni collisions at 1.93~AGeV. 
Data are shown by the dots. The shaded area shows an estimate for systematic 
errors. The solid lines show BUU calculations without in-medium effects. 
The dotted (dashed) lines correspond to $\rm U_{K^{+}} = 30~MeV$ and 
$\rm U_{K^{-}} = -120~MeV$ (-70 MeV).} 
\label{ratio} 
\end{figure}

The FOPI collaboration has also investigated the $\rm K^{-}$/$\rm K^{+}$ 
ratio as a function of rapidity in Ru+Ru collisions at 1.69~AGeV and 
Ni+Ni collisions at 1.93~AGeV~\cite{wis00}. Results are presented on 
figure~\ref{ratio} for central collisions. Experimental data are shown by 
the dots. The ratio increases from target rapidity to mid-rapidity reflecting 
the fact that the kaon production in a dense medium is reduced due to the 
repulsive potential whereas the antikaon production is enhanced due to the 
presence of  an attractive antikaon-nucleon potential.\\ 
Data are compared to the predictions of the BUU model without in-medium 
effects (solid line) and with two different $\rm K^{-}$ potentials (dotted 
and dashed lines). The version without in-medium modifications of kaon 
masses fails to reproduce the trend observed in the data. The calculations 
including in-medium potential show the same trend as the experimental data, 
which seem to favor the smaller antikaon potential. This corresponds to a 
$\rm K^{-}$ mass reduction by 12$\%$ at normal nuclear matter density.

\section{Strangeness propagation}

\begin{figure}[hbt]
\vspace{1cm}
\centering\mbox{\epsfig{file=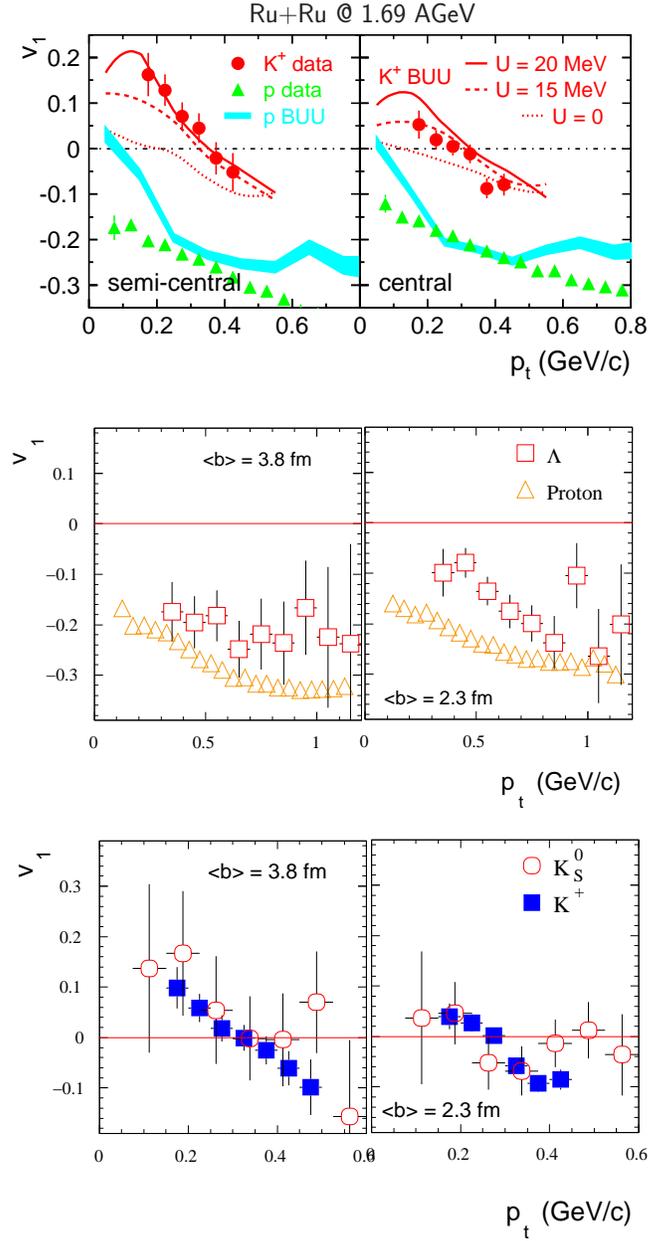, width=0.54\textwidth}}
\vspace{2mm}
\caption{First Fourier coefficient as a function of transverse momentum 
in Ru+Ru semi-central (left) and central (right) collisions at 1.69~AGeV.
Upper panel: $\rm K^{+}$ (dots) and protons (triangles) compared to the 
predictions of the BUU model (shaded area: protons, dotted line: without 
potential, dashed line: U=15 MeV, solid line: U=20 MeV). Middle panel: 
$\Lambda$ (squares) and protons (triangles). Lower panel: $\rm K^{+}$ 
(squares) and $\rm K^{0}$ (circles).} 
\label{flowkp} 
\end{figure}

The differential sideward flow of strange particles is presented on 
figure~\ref{flowkp} for Ru+Ru central and semi-central collisions at 
1.69~AGeV in terms of the first Fourier coefficient ($\rm v_{1} = <cos(\phi)>$ 
where $\phi$ is the angle between the particle and the reaction plane) as a 
function of the transverse momentum. The upper panel shows the results 
for $\rm K^{+}$ (dots) and protons (triangles)~\cite{cro00}. The kaon 
sideward flow varies with the transverse momentum from positive values 
of $\rm v_{1}$ to negative values. The shaded area shows the BUU calculations 
for protons. The dotted line shows BUU calculations without in-medium 
modifications. The solid and dashed lines show the calculations for two 
potentials. The kaon data clearly favor the versions with in-medium 
potential. Note that the sensitivity to in-medium effects is the largest at 
low transverse momenta.

\newpage

The middle panel shows results for $\Lambda$ (open squares) compared to 
protons (open triangles)~\cite{kut00}. Although $\Lambda$ are produced 
with the same mechanism as $\rm K^{+}$, their flow pattern is very 
different. Their flow follows the one of the protons although the magnitude 
is a little lower.\\
Results for $\rm K^{0}$ (open circles)~\cite{kut00} are compared to 
$\rm K^{+}$ flow (squares) in the lower panel. Within error bars, neutral 
kaon flow is compatible with charged kaon flow which evidences a very weak 
effect of the Coulomb potential.

\section{Summary}

The FOPI collaboration has measured strangeness production and propagation 
in nuclear matter at SIS energies. Results on $\rm K^{+}$ and $\rm K^{0}$ 
rapidity distributions as well as $\rm K^{-}$/$\rm K^{+}$ ratio in Ru+Ru 
collisions at 1.69~AGeV have been presented. The differential sideward flow 
of $\rm K^{+}$, $\rm K^{0}$ and $\rm \Lambda$ has been measured in Ru+Ru 
collisions at 1.69~AGeV and Ni+Ni collisions at 1.93~AGeV.\\
Experimental data are compared to the predictions of transport model 
calculations (BUU and IQMD). Although the set of available data is quite 
large, it is still not possible to draw definitive conclusions on the 
presence of in-medium modifications of hadron properties. Some observables, 
like rapidity distributions, seem to favor a scenario where kaon masses 
are not modified in-medium whereas others, like the flow, favor calculations 
including in-medium modification of hadron masses. It appears that none of 
the models used in this work is able to reproduce all observables with one 
set of parameters. \\


\section*{References}
\begin{thebibliography}{9}

\bibitem{raf82}
J. Rafelski and B. M\"uller 1982 {\it Phys. Rev. Lett.} {\bf 48} 1066
\bibitem{bro91}
G.E. Brown and M. Rho 1991 {\it Phys. Rev. Lett.} {\bf 66} 2720
\bibitem{cro00}
P. Crochet {\it et al} (FOPI collaboration) 2000 {\it Phys. Lett.} {\bf 486} 6
\bibitem{gob93}
A. Gobbi {\it et al} (FOPI collaboration) 1993 {\it Nucl. Inst. Meth.} 
{\bf A324} 156
\bibitem{stu01}
C. Sturm {\it et al} (KaoS collaboration) 2001 {\it Phys. Rev. Lett.} 
{\bf 86} 39
\bibitem{dev01}
A. Devismes 2001, PhD Thesis, University of Technology, Darmstadt, Germany
\bibitem{bes97}
D. Best {\it et al} (FOPI collaboration) 1997 {\it Nucl. Phys.} {\bf A625} 307 
\bibitem{cas99}
W. Cassing and E.L. Bratkovskaya 1999 {\it Phys. Rep.} {\bf 308}
\bibitem{har92}
Ch. Hartnack 1992, PhD Thesis, University of Frankfurt, Germany
\bibitem{har01}
Ch. Hartnack 2001, these proceedings
\bibitem{wis00}
K. Wi\'sniewski {\it et al} (FOPI collaboration) 2000 {\it Eur. Phys. J.}
 {\bf A9} 515
\bibitem{kut00}
R. Kutsche 2000, PhD Thesis, University of Technology, Darmstadt, Germany
\end{thebibliography}
\end{document}